\documentclass[twocolumn,aps,pre,footinbib,floatfix,preprintnumbers,amsmath,amssymb,superscriptaddress,10pt,colorlinks,a4paper]{revtex4-1}

\usepackage{ulem}
\renewcommand{\emph}{\textit}

\usepackage{selinput}
\usepackage[left=1.6cm,right=1.6cm,bottom=1.7cm,top=1.7cm]{geometry}
\usepackage[usenames,dvipsnames]{color}
\usepackage{natbib}
\usepackage{dsfont}
\usepackage{xr}
\usepackage{tensor}
\usepackage[T1]{fontenc}
\usepackage[english]{babel}
\usepackage{lmodern}
\usepackage{textcomp}		
\usepackage{textgreek}		
\usepackage{enumerate}
\usepackage{amsmath}		
\usepackage{amssymb}	
\usepackage{amsfonts}
\usepackage{mathrsfs}
\usepackage{braket}
\usepackage{mathtools}
\usepackage{siunitx}	
\usepackage{bm}		
\usepackage[version=4]{mhchem}		
\usepackage{graphicx}	
\usepackage[export]{adjustbox}
\usepackage{lettrine}
\usepackage[english]{varioref}
\usepackage[urlcolor=RoyalBlue,citecolor=RoyalBlue,linkcolor=Black]{hyperref}

\DeclarePairedDelimiter{\abs}{\lvert}{\rvert}

\newcommand{\mean}[1]{\left < #1 \right >}

\renewcommand{\epsilon}{\varepsilon}
\renewcommand{\theta}{\vartheta}
\renewcommand{\phi}{\varphi}

\newcommand{\dicty}{\textit{D.~discoideum}}

\renewcommand{\vec}[1]{\mathbf{ #1 }}

\makeatletter
\renewcommand\section{%
  \@startsection
    {section}%
    {2}%
    {-10pt}%
    {-1.5em}%
    {3pt}%
    {\sffamily\large\bfseries}%
}%
\makeatother

\makeatletter
\renewcommand\paragraph{%
  \@startsection
    {paragraph}%
    {3}%
    {-\parindent}%
    {-1.8em}%
    {0.5pt}%
    {\sffamily\normalsize\bfseries}%
}%
\makeatother

\makeatletter
\newcommand\metparagraph{%
  \@startsection
    {paragraph}%
    {3}%
    {\parindent}%
    {0em}%
    {-0.6em}%
    {\newline \sffamily\small\bfseries}%
}%
\makeatother

\renewcommand{\figurename}{Figure}

\makeatletter
\renewcommand{\fnum@figure}{\small{\sffamily{\textbf{\figurename~\thefigure}}}\normalfont}
\renewcommand*{\@caption@fignum@sep}{ \small{$\boldsymbol{|}$} }
\makeatother

\newcommand{\fighead}[1]{\small{\sffamily{\textbf{#1}}}\normalfont}

\newcommand{\fl}[1]{\small{\textbf{\sffamily{#1}}}\normalfont}

\begin{document}

\title{\large \sffamily{Biohybrid active matter~--~the emergent properties of cell-mediated microtransport}} 

\author{Valentino Lepro}
\thanks{These two authors contributed equally. }
\affiliation{Institute of Physics and Astronomy, University of Potsdam, Karl-Liebknecht Stra{\ss}e 24/25, 14476 Potsdam, Germany}
\affiliation{Max Planck Institute of Colloids and Interfaces, Am M\"uhlenberg 1, 14476 Potsdam, Germany}

\author{Robert Gro{\ss}mann}
\thanks{These two authors contributed equally. }
\affiliation{Institute of Physics and Astronomy, University of Potsdam, Karl-Liebknecht Stra{\ss}e 24/25, 14476 Potsdam, Germany}

\author{Oliver Nagel}
\affiliation{Institute of Physics and Astronomy, University of Potsdam, Karl-Liebknecht Stra{\ss}e 24/25, 14476 Potsdam, Germany}

\author{Setareh~Sharifi~Panah}
\affiliation{Institute of Physics and Astronomy, University of Potsdam, Karl-Liebknecht Stra{\ss}e 24/25, 14476 Potsdam, Germany}

\author{Stefan Klumpp}
\affiliation{Max Planck Institute of Colloids and Interfaces, Am M\"uhlenberg 1, 14476 Potsdam, Germany}
\affiliation{Institute for the Dynamics of Complex Systems, Georg August University of G\"ottingen, Friedrich-Hund-Platz 1, 37077 G\"ottingen, Germany}

\author{Reinhard Lipowsky}
\affiliation{Max Planck Institute of Colloids and Interfaces, Am M\"uhlenberg 1, 14476 Potsdam, Germany}

\author{Carsten Beta}
\email{beta@uni-potsdam.de}
\affiliation{Institute of Physics and Astronomy, University of Potsdam, Karl-Liebknecht Stra{\ss}e 24/25, 14476 Potsdam, Germany}

\begin{abstract}

As society paves its way towards device miniaturization and precision medicine, micro-scale actuation and guided transport become increasingly prominent research fields with high impact in both technological and clinical contexts. 
In order to accomplish directed motion of micron-sized objects towards specific target sites, active biohybrid transport systems, such as motile living cells that act as smart biochemically-powered micro-carriers, have been suggested as an alternative to synthetic micro-robots. 
Inspired by the motility of leukocytes, we propose the amoeboid crawling of eukaryotic cells as a promising mechanism for transport of micron-sized cargoes and present an in-depth study of this novel type of composite active matter.
Its transport properties result from the interactions of an active element~(cell) and a passive one~(cargo) and reveal an optimal cargo size that enhances the locomotion of the load-carrying cells, even exceeding their motility in the absence of cargo.
The experimental findings are rationalized in terms of a biohybrid active matter theory that explains the emergent cell-cargo dynamics and enables us to derive the long-time transport properties of amoeboid micro-carries.  
As amoeboid locomotion is commonly observed for mammalian cells such as leukocytes, our results lay the foundations for the study of transport performance of other medically relevant cell types and for extending our findings to more advanced transport tasks in complex environments, such as tissues. 

\end{abstract}

\maketitle

The targeted delivery of micron-sized objects, such as drug-releasing microparticles or nanoelectronic biosensors, is one of the prime challenges in modern medical technology.
Ideally, future solutions will rely on mechanisms of self-propulsion, allowing micron-sized cargoes to actively navigate through complex and crowded environments such as human tissue~\cite{sitti2015biomedical,xu_self-propelled_2020}.
To achieve this goal, much effort has been devoted to the design of bio-inspired synthetic micro-robots~\cite{bayley_stochastic_2001,sanchez_nanorobots_2009,patino_miniaturized_2016,paoluzzi_shape_2016,joseph_chemotactic_2017,hess_nonequilibrium_2017,tang_enhancing_2018,soto_medical_2020}.
However, several technical difficulties limit this approach, such as questions of power supply, biocompatibility, and efficient steering, especially in complex environments~\cite{sitti2009voyage,carlsen2014bio,abdelmohsen2014micro,sitti2015biomedical,wang2013small,hwang2011electro,ricotti2017biohybrid,yasa2020elucidating,zhang_cooperative_2021}.
Rather than reinventing nature's sophisticated machines, an alternative paradigm to address this challenge is to take advantage of them directly:~by loading micro-cargoes onto motile cells, their innate migratory abilities can be exploited to achieve directed transport in a biohybrid approach~\cite{ricotti2017biohybrid,sitti2015biomedical,carlsen2014bio,alapan2019microrobotics,sun_biohybrid_2020,pacheco_functional_2021}.
While this principle has been successfully demonstrated with the help of self-propelled bacterial swimmers, the physical properties and, particularly, the transport capacities of the most common modes of eukaryotic locomotion, such as amoeboid crawling of leukocytes during an inflammatory response~\cite{friedl2008interstitial,wolf2003amoeboid,titus2017evolutionary}, have remained mostly unexplored and were only recently exemplified in a few cases~\cite{anselmo2015monocyte,shao_chemotaxis-guided_2017,xue_neutrophil-mediated_2017}.

At a theoretical level, the physical properties of actively moving entities, nowadays referred to as active matter, is one of the most rapidly evolving, newly established subfields of physics that connects fundamental questions of nonequilibrium thermodynamics and statistical mechanics with current challenges and open questions in our understanding of complex biological systems~\cite{marchetti2013soft,chate2020dry,bar2020self}.
While early active matter research has mostly been driven by fundamental theoretical questions~\cite{vicsek1995novel,toner1995long,chate2020dry}, the field is now increasingly focusing on specific experimental model systems, studying, for example, the formation of emergent patterns in motor-driven assemblies of cytoskeletal filaments~\cite{huber_emergence_2018}, the movement of bacterial swimmers~\cite{gronot2021more}, or collective motion~\cite{vicsek2021collective}. 
However, the potential of cargo-carrying active particles for guided transport has been largely ignored and has only recently received increasing interest~\cite{vuijk_chemotaxis_2021,jin_collective_2021}.
A general theoretical framework for biohybrid active matter that is tested in the light of experimental observations has not been established so far.

\begin{figure*}[thp!]
\centering
\includegraphics[width=0.8\textwidth]{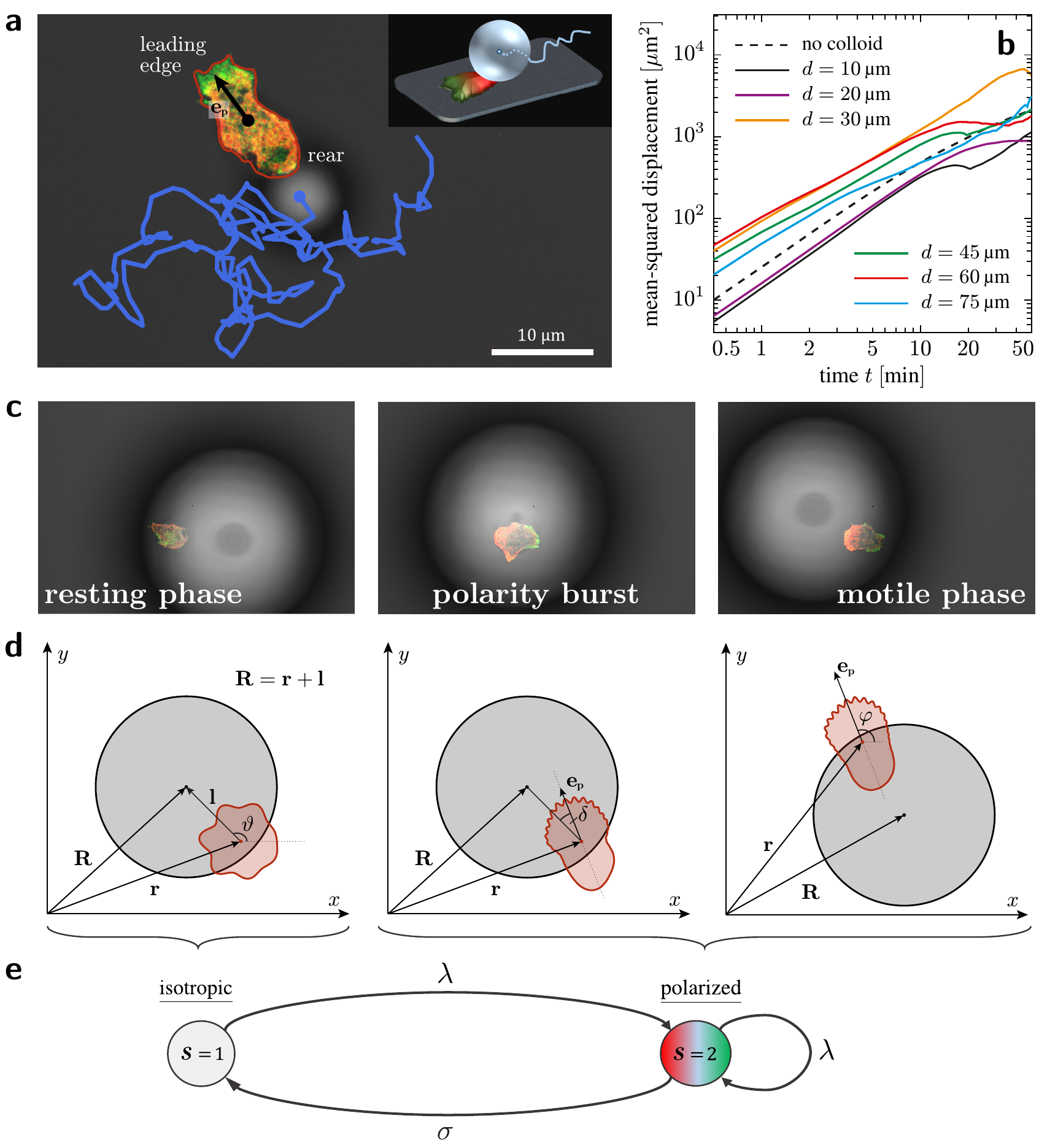}
\caption{\fighead{Phenomenology of the cell-cargo truck and its dynamics.} Panel~\fl{a} shows cell and cargo as observed experimentally together with an exemplary trajectory of the cargo. The snapshot shows a polarized cell as indicated by a black arrow, where F-actin is color-coded in green and myosin II in red. An illustration of the three-dimensional spatial configuration of cell and cargo is presented in the inset. The mean time-averaged mean-squared displacement~(MSD) of cargoes, transported by the cell, is shown in panel~\fl{b} for different cargo sizes; note that less data points for the time-averaging procedure are available for large time lags, which is the reason why the curves become noisy in the long-time limit~(cf.~Methods for additional technical details). Notably, the transport efficiency is non-monotonic in particle size and reaches a maximum for intermediate sizes. Moreover, the MSD of cells that are loaded with a particle can exceed the MSD of unloaded cells~(dashed line in panel~\fl{b}). The three panels in~\fl{c} indicate the typical phases of cellular truck dynamics. In~\fl{d}, key geometric quantities are introduced:~the position of cell~$\vec{r}$ and cargo~$\vec{R}$, the distance vector~$\vec{l}=\vec{R}-\vec{r}$ with its polar angle $\theta$, the cell polarization vector~$\vec{e_p}$ along with the cell polarization angle~$\varphi$, and the angle~$\delta$ indicating the direction of initial cell polarization compared to the cell-cargo axis, given by the distance vector~$\vec{l}$. The sequence of cell polarization and depolarization is illustrated in panel~\fl{e} together with the corresponding rates of occurrence~($\lambda$ and $\sigma$, respectively). For movies, see~SM~\cite{SM}. }
\label{fig:1}
\end{figure*}

In this work, we combine live cell experiments with theoretical modeling to study the fundamental principles of active biohybrid transport driven by adherent eukaryotic cells.
The active biohybrid is composed of an amoeboid cell as an active element and a polystyrene bead ranging in diameter from $10 \;\! \mbox{\textmugreek m}$ to $100 \;\! \mbox{\textmugreek m}$ as cargo.
We rely on cells of the social amoeba {\it Dictyostelium discoideum} ({\dicty}), a well-established model organism for actin-driven motility of eukaryotic cells that shares many similarities with neutrophils~\cite{friedl2001amoeboid,artemenko2014moving}. 
Given the highly non-specific adhesion of our model organism~\cite{loomis2012innate}, the binding of micro-cargo to the cell membrane does not require any surface functionalization, whereas this may be required for specific applications with other cell types.
The physical link between cargo and carrier is established spontaneously; as a cell gets in contact with a particle, the cargo adheres to the cell membrane and is then subjected to forces exerted by the cell. 
At a qualitative level, biohybrid transport of micron-sized cargoes has recently been demonstrated for this cell type~\cite{nagel2019harnessing}.
We now present an in-depth experimental study of this composite cell-cargo system that serves as a basis to establish a general theoretical framework for biohybrid active matter.
We show how the cell-cargo interactions, due to the mechano-responsiveness of amoeboid cells, shape the long-time transport properties.
In particular, our analysis reveals a non-monotonic dependence of the diffusivity on cargo size, thus demonstrating the existence of an optimal cargo size which maximizes the transport efficiency, even exceeding the spreading of single cells in the absence of cargo.


\paragraph*{Cell-cargo system shows multiple transport phases}

\begin{figure*}[htp!]
\centering
\includegraphics[width=0.9\textwidth]{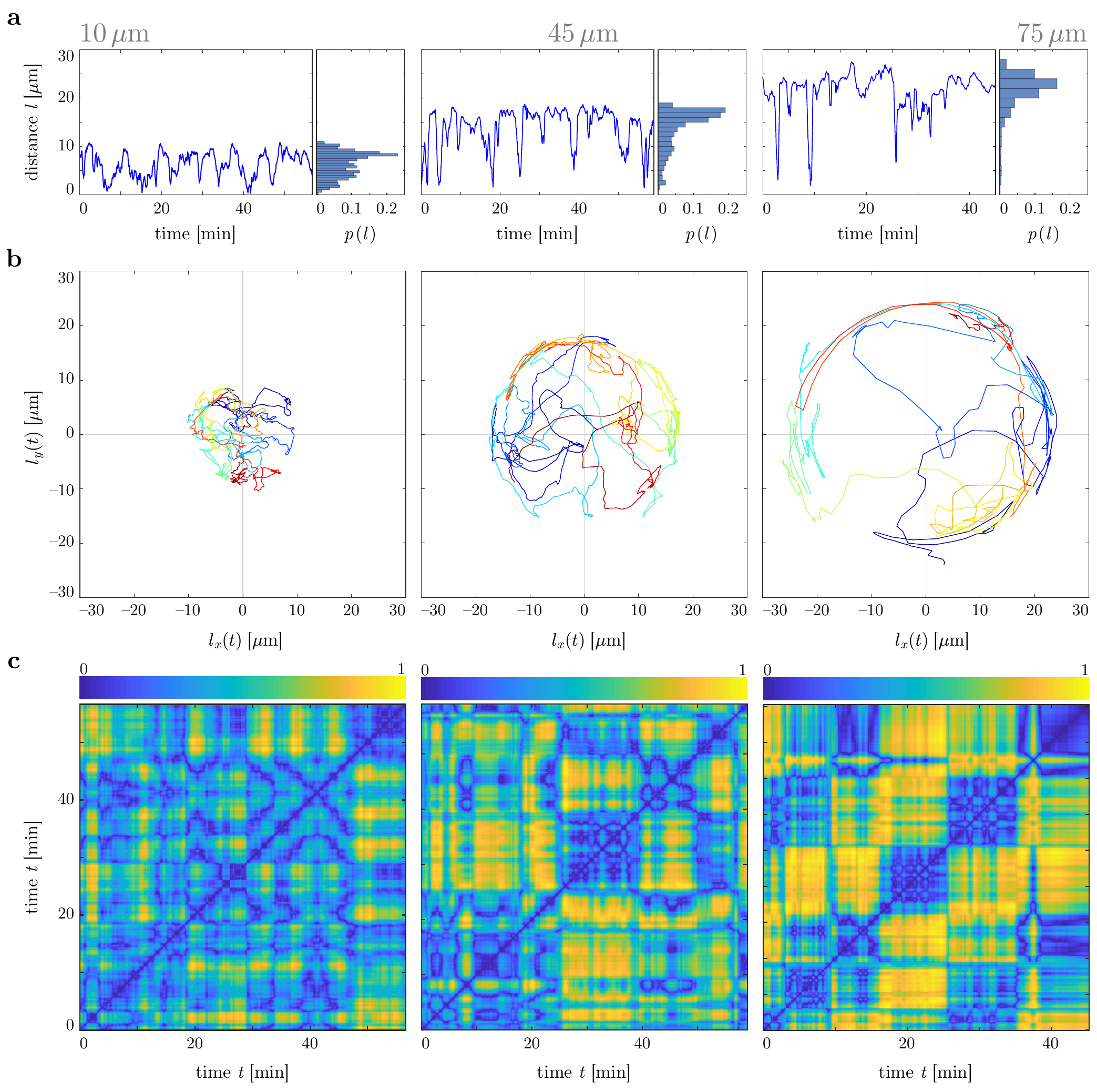}
\caption{\fighead{Relative dynamics of cell and cargo.} In~\fl{a}, time series of the distance~$l(t)=\abs{\vec{R}(t)-\vec{r}(t)}$ between cell and cargo are shown together with the corresponding histograms. The dynamics of the cargo in the frame of reference of the cell is plotted in~\fl{b}, where time is color coded. Panels~\fl{c} display color coded recurrence plots~$I(t_i,t_j)=[\Delta_{ij} - \mbox{min}(\Delta_{ij})] / [\mbox{max}(\Delta_{ij}) - \mbox{min}(\Delta_{ij})]$, where $\Delta_{ij} = \abs{\vec{l}(t_j)-\vec{l}(t_i)}$ and~$\vec{l}(t)=\vec{R}(t)-\vec{r}(t)$. Columns correspond to three different particle diameters:~$d=10\!\;\mbox{\textmugreek m}$~(left),~$d=45\!\;\mbox{\textmugreek m}$~(middle), and~$d=75\!\;\mbox{\textmugreek m}$~(right).}
\label{fig:2}
\end{figure*}

A general view of the biohybrid cell-cargo system, referred to hereafter as \textit{cellular truck} or \textit{truck}, is given in Fig.~\ref{fig:1}a along with a schematic in the inset, depicting its typical conformation: the particle resides on top of the cell, typically towards its rear edge, while the cell moves forward, thereby pulling the cargo~(see the Supplemental Material~(SM) for movies~\cite{SM}). 
Tracking of the colloid position reveals that the mean-squared displacements of the cargo particle, shown in Fig.~\ref{fig:1}b, is non-monotonic in particle size:~notably, for small cargo particles the transport efficiency increases with increasing cargo size until an optimal, intermediate cargo di\-a\-me\-ter is reached, which maximizes the transport efficiency of the cellular truck.
To elucidate the mechanism that leads to these counter-intuitive transport properties of cellular trucks is one of the central aims of this study.

We performed dual-color fluorescence imaging experiments, where the motion of cell and cargo are resolved, while at the same time visualising the polarization of the cytoskeleton.
For this we used a \textit{D.~discoideum} cell line that expressed fluorescent labels of F-actin and myosin~II, known to localize at the leading edge and at the rear of a polarized cell, respectively~\cite{rappel2017mechanisms,dalous2008reversal}. 
In the following, we denote cell polarization by the unit vector~$\vec{e_p}$, and the centers of cell and cargo by the two dimensional position vectors~$\vec{r}$ and~$\vec{R}$, respectively, defined in the focal plane of imaging, which is parallel to the substrate surface~(cf.~Methods). 
This experimental setting enabled us to study the cargo dynamics with respect to the cell in terms of the distance vector
\begin{equation}
    \vec{l}(t) = \vec{R}(t) - \vec{r}(t), 
\end{equation}
cf.~Fig.~\ref{fig:1} for an illustration.

From our microscopy recordings, we identified two distinct transport phases that robustly emerged for all tested particle sizes, cf.~Fig.~\ref{fig:1}c-e.
During \emph{resting phases}, cell and particle move around each other, keeping their distance~$l(t)=\abs{\vec{l}(t)}$ approximately constant without significant net displacement of the whole truck.
In this phase, the cell shows no or only short-lived cytoskeletal polarization. 
Trucks can dwell in the resting phase for up to several tens of minutes. 
In addition, we observed intermittent \emph{polarity bursts}, where the cell suddenly reorients and polarizes towards the cargo, crawling underneath and at the same time pulling the cargo towards its newly defined rear. 
This process usually takes a few minutes; we call its rate of occurrence~$\lambda$~(polarization rate).
After such transitions, a \emph{motile phase} follows, where cell polarization is maintained, causing the cell to keep moving and pulling the cargo forward.
This phase is associated with significant displacement of the cellular truck. 
Once the cell spontaneously depolarizes, typically after a polarity lifetime~$\tau_p = \sigma^{-1}$ of a few minutes, the persistent run ends and the truck returns to the resting phase, waiting for another polarization event to be triggered. 
The cell may also repolarize towards the cargo while being in the motile phase; in such cases, the truck ends its current run and directly enters the next polarity burst, see Fig.~\ref{fig:1}e for a schematic representation of the sequence of cell polarization and depolarization together with the corresponding rates of occurrence. 
With respect to the cell, we may thus distinguish two states $s$, an isotropic ($s=1$) and a polarized state ($s=2$), where the latter incorporates both the polarity burst and the subsequent motile phase.

\paragraph*{Cell-cargo distance displays recurrent dynamics}

The cyclic change between these transport states gives rise to a recurrent yet stochastic dynamics for the distance~$l(t) = \abs{\vec{l}(t)}$ between cell and particle as illustrated for three different particle sizes in Fig.~\ref{fig:2}a.
The distance fluctuates around a preferred value~$l_0$, interrupted by repeated abrupt decays that correspond to the polarity-induced crossing events, when the cell moves underneath the particle, shifting it towards its other side.
The value of the preferred distance $l_0$ increases with increasing cargo size.
Histograms of the time-series of the distance consistently display a peaked distribution with a negative skew due to the crossing events.
Note that we also observed pronounced differences in the polarization rate~$\lambda$ for different cargo sizes:~the average time interval~$\tau = \lambda^{-1}$ between polarity bursts increases with the particle diameter~$d$.

A more detailed look at the relative cell-cargo dynamics is presented in Fig.~\ref{fig:2}b.
Here, the frame of reference is aligned with the center of the cell~$\vec{r}(t)$, and the relative cargo trajectory~$\vec{l}(t)=\vec{R}(t)-\vec{r}(t)$ is plotted using a color scale to encode time.
Typically, the cell pulls the cargo randomly around itself, keeping on average a characteristic distance $l_0$.
From time to time, however, the cell moves towards the particle center, crossing to the other side. 
The larger the particle, the more prominent this motion pattern emerges, see the trajectories of the relative motion of cells loaded with $45 \!\; \mbox{\textmugreek m}$ and $75 \!\; \mbox{\textmugreek m}$ particles.

To obtain a more quantitative representation of the cell-cargo dynamics, we generated recurrence plots~\cite{eckmann_recurrence_1987,marwan2007recurrence} from the time evolution of the distance vector~$\vec{l}(t)$.
Recurrence plots graphically represent to which degree a dynamical system revisits similar areas in phase space at times~$t_i$ and~$t_j$ by means of a matrix~$I(t_i,t_j)$. 
In Fig.~\ref{fig:2}c, examples of recurrence plots are shown, where~$I(t_i,t_j)$ is color-coded.
They correspond to the relative motion of the three examples shown in panels~a and~b.
The recurrence plots display a distinctive checkerboard pattern:~in our context, patches correspond to resting and motile phases, separated by sharp boundaries that are related to the randomly occurring polarity bursts.
These phases become increasingly pronounced for larger cargoes. 
Furthermore, the polarization rate decreases for larger particles, which is reflected by larger patches in the recurrence plots. 
Taken together, the analysis of recurrence plots suggests an intermittent relative dynamics, where dwelling at a certain preferred cell-particle distance~$l(t) \approx l_0$ alternates with stochastically occurring sudden episodes of relative motion, during which the cell polarizes towards the cargo, moves underneath it, and the cargo is simultaneously pulled from one side of the cell to the other.

\begin{figure}[t]
\centering
\includegraphics[width=\columnwidth]{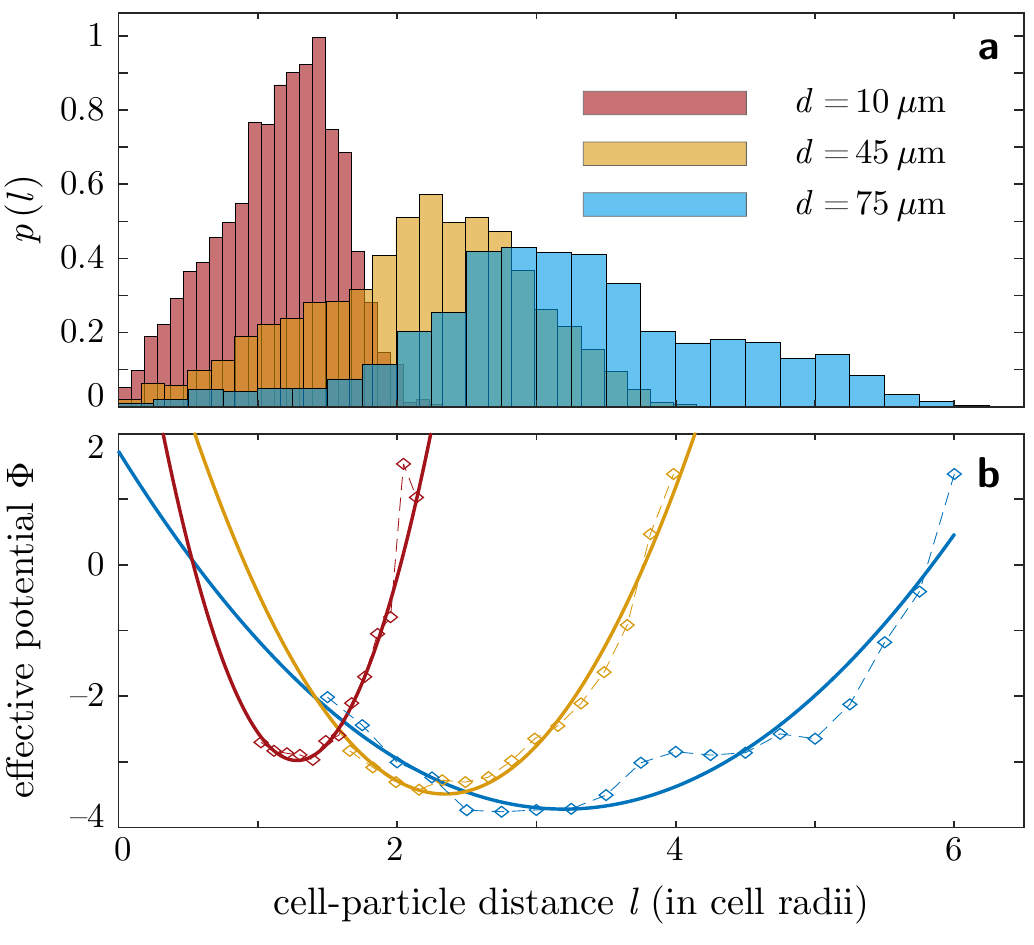}
\caption{\fighead{Population-averaged histogram of cell-cargo distance and their effective interaction potential.} In panel~\fl{a}, population averaged histograms of the cell-particle distance, measured in multiples of an effective cell radius~$r_c$ to account for cell size variability, are shown~(cf.~Methods for calculation of $r_c$). Population averages were taken over ten independent experimental realizations for each cargo size. Based on these histograms, effective potentials were inferred~(panel~\fl{b}) via the Boltzmann formula~$p(l)\sim e^{-\Phi(l)}$. The peak of the histogram relates to an effective harmonic interaction potential; fits using quadratic functions are shown by solid lines in~\fl{b}. Only those data points which are included into the fit were plotted; for~$l\rightarrow 0$, the potential is expected to be anharmonic due to the polarization dynamics~(Fig.~\ref{fig:1}c), which is why those data points were excluded from fitting.  }
\label{fig:3}
\end{figure}

Additionally, we calculated population-averaged histograms of the distance~$l(t)$ by binning several time series for a given particle size~(see Fig.~\ref{fig:3}a). 
The emerging ensemble-averaged distributions retained a peaked shape as for the single truck. 
With increasing particle diameter~$d$, the peak position shifts towards larger values and the histograms become wider.
Given these distributions, we inferred an effective cell-cargo interaction potential~$\Phi(l)$ in the vicinity of the peak by Boltzmann inversion of the histogram~$p(l)$ via
\begin{align}
    p(l) \sim e^{- \Phi(l)} 
\end{align}
relating the probability distribution of a physical quantity to an effective energy landscape~\cite{coughlan2017effective}. 
As shown in Fig.~\ref{fig:3}b, the inferred potentials are harmonic close to the peak position. 
Together with the recurrence plot analysis, this suggests that the cell-cargo dynamics can be described as an elastic interaction, recurrently perturbed by polarity bursts in the course of which the cell moves underneath the cargo, pulling it onto its other side.

\paragraph*{An active particle model of cellular truck dynamics}

Based on the experimental observations, we propose a phenomenological model that mimics the dynamics of a cellular truck. 
The cell is the actual active element~--~any displacement of cell and particle results from active forces exerted by the cell onto the cargo or the substrate.
In contrast, thermal diffusion of the cargo is negligible for the particle sizes considered. 
The model relies on the observation that the cell can adopt two states at random:~an \textit{isotropic state}~($s=1$) and a \textit{polarized state}~($s=2$), see Fig.~\ref{fig:1}.

In the isotropic state~($s=1$), the cell moves diffusively with an effective diffusion coefficient~$D_i$. 
The particle, adherent to the cell, is pulled around it randomly; the cell-cargo distance fluctuates around a characteristic value~$l_0$ which depends on the size of the particle. 
The key mechanism that determines the dynamics of the truck is the cell's tendency to polarize towards the cargo. 
In the polarized state~($s=2$), the cell exerts net forces on its environment, thereby~(i) enabling persistent motion at a non-vanishing speed~$v_0$ that we assume to be constant for simplicity and (ii) pulling the particle onto its back with respect to the cell polarization vector 
\begin{align}
\label{eqn:rg:mod1}
	\vec{e_p} \! \left[ \varphi (t) \right] = \begin{pmatrix} \cos \varphi(t) \\ \sin \varphi(t) \end{pmatrix} 
\end{align}
which is parametrized in terms of the polarization angle~$\varphi$ and defined to have unit length~(cf.~Fig.~\ref{fig:1}).

Mathematically, the spatial dynamics of the cell's center~$\vec{r}(t)$ is described by the following equations: 
\begin{subequations}
\label{eqn:mot-cell}
\begin{align}
	\dot{\vec{r}}(t) &= v(s) \!\: \vec{e_p} \! \left[ \varphi (t) \right] + \!\!\: \sqrt{2D(s)} \:\! \boldsymbol{\xi}_{\vec{r}} \!\!\; (t),  \label{eqn:mot-cella} \\
	\dot{\varphi}(t) &= \sqrt{2 D_{\varphi}} \:\! \xi_{\varphi} \!\!\; (t). \label{eqn:mot-cellb}
\end{align}
\end{subequations}
The speed~$v(s)$ and the effective diffusion coefficient~$D(s)$ are state dependent:~$v(1)=0$ in the isotropic state, whereas~$v(2) = v_0$ in the polarized state; furthermore, we denote~$D(1)= D_i$ and~$D(2) = D_p$, respectively. 
The terms~$\boldsymbol{\xi}_{\boldsymbol{r}}$ and $\xi_{\varphi}$ represent independent Gaussian white noise processes with zero mean and temporal~$\delta$-correlations~\cite{gardiner_stochastic_2009}.

\begin{figure*}[t]
\centering
\includegraphics[width=\textwidth]{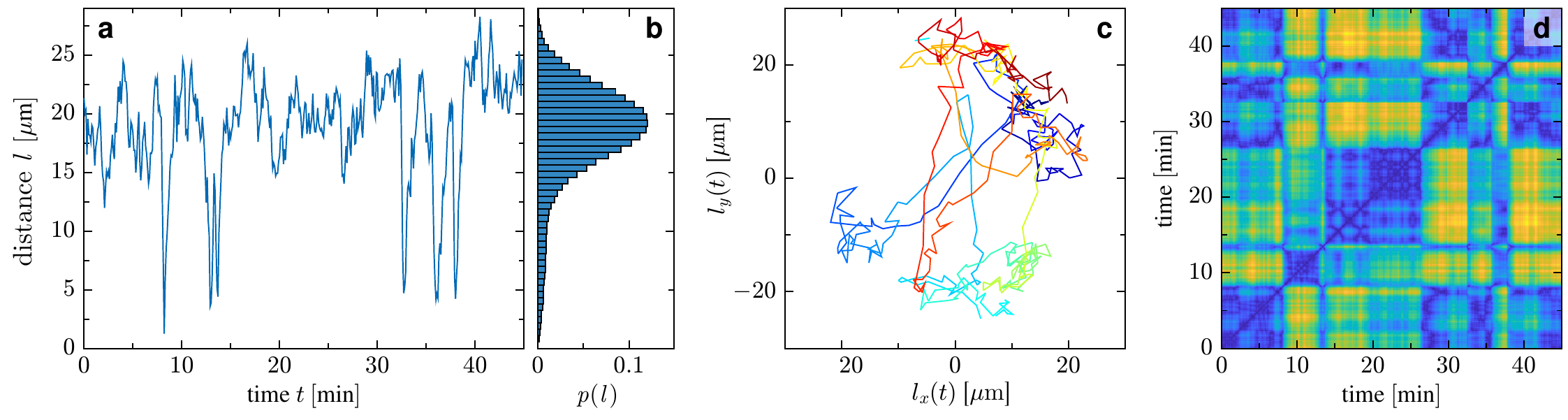}
\caption{\fighead{Predictions of the active particle model for the cell-cargo dynamics.} The representation of simulation results corresponds to the experimental results displayed in Fig.~\ref{fig:2}: (\fl{a})~time series of the distance of the centers of mass of cell and cargo together with the corresponding histogram~(panel~\fl{b}); (\fl{c})~trajectory of the cargo in the frame of reference of the cell; (\fl{d})~recurrence plot~$I(t_i,t_j)$ of the distance vector~$\vec{l}(t)=\vec{R}(t)-\vec{r}(t)$, cf.~Fig.~\ref{fig:2}. All features of the experimentally observed cellular truck dynamics are qualitatively reproduced by the model. Parameter values:~$\sigma=0.1 \!\; \mbox{min}^{-1}$, $\lambda=0.139 \!\; \mbox{min}^{-1}$, $\alpha=2 \!\; \mbox{min}^{-1}$, $D_{R} = 13 \;\! \mbox{\textmugreek m}^2/\mbox{min}$, $l_0 = 18 \;\! \mbox{\textmugreek m}$, $v_0 = 6 \;\! \mbox{\textmugreek m}/\mbox{min}$, $D_{\varphi} = 0.1 \;\! \mbox{min}^{-1}$, $D_i = 1 \;\! \mbox{\textmugreek m}^2/\mbox{min}$, $D_p = 1 \;\! \mbox{\textmugreek m}^2/\mbox{min}$, $\delta \sim \mathcal{N}(0,0.2)$, numerical time step~$\Delta t = 0.01 \!\; \mbox{min}$~(stochastic Euler method~\cite{gardiner_stochastic_2009}). } 
\label{fig:4}
\end{figure*}

The position of the particle~$\vec{R}(t)$, adherent to the cell, is described by a linear interaction term~(cf.~Fig.~\ref{fig:3}) and unbiased Gaussian white noise $\boldsymbol{\xi}_{\boldsymbol{R}}$ that accounts for non-thermal displacement due to undirected forces exerted on the particle by the cell: 
\begin{align}
	\label{eqn:mot-coll2}
	\dot{\vec{R}}(t) & = - \alpha \Big( \vec{R} - \vec{R}_0(s) \Big ) \!\!\: + \!\!\: \sqrt{2D_{R}} \, \boldsymbol{\xi}_{\boldsymbol{R}} \!\!\; (t). 
\end{align}
In the passive state~($s=1$), the distance~$l = \abs{\vec{R} - \vec{r}}$ between cell and particle fluctuates around a characteristic value~$l_0$, while the particle can circle around the cell. 
The ratio~$D_R/\alpha$ of the noise amplitude~$D_R$ and the spring constant~$\alpha$ determine the strength of fluctuations of~$l$ and, thereby, the width of the histogram of the cell-particle distance~$l$. 
Moreover,~$1/\alpha$ is the characteristic timescale of the cell-particle dynamics. 
In the polarized state, the particle moves to the back of the cell with respect to its polarization~$\vec{e_p}$. 
This is modeled by a state-dependent minimum~$\vec{R}_0(s)$ of the interaction potential in Eq.~\eqref{eqn:mot-coll2}: 
\begin{align}
	\vec{R}_0(s) = \vec{r} + l_0 \!\!\: \cdot \!\!\: 
	\begin{cases}
		  \vec{l} / \abs{\vec{l}} , & s = 1, \\ 
		  - \vec{e_p}[\varphi], & s = 2. 
	\end{cases} 
\end{align}
Lastly, we describe transitions between the isotropic and polarized states~($s=1$ and~$s=2$, respectively) as Poisson processes with the transition rates~$\{ \lambda,\sigma \}$, see~Figs.~\ref{fig:1}d,e.
In the isotropic state, the cell does not exhibit any polarization, cf.~Eq.~\eqref{eqn:mot-cella} with $v(1) = 0$. 
At each polarity burst, occurring at a rate~$\lambda$, the vector~$\vec{e_p}$ is initialized anew and now affects the dynamics of the cell as $v(2) = v_0$ in Eq.~\eqref{eqn:mot-cella}. 
Due to the mechanical stimulus of the cargo, the cell tends to polarize along the cell-cargo axis~$\vec{l}=l \! \left ( \cos \theta, \sin \theta \right )$ on average, i.e.~towards the current position of the cargo, plus some angular variability~$\delta$. 
A polarity burst is therefore mathematically described by resetting the polarization angle~$\varphi \rightarrow \theta + \delta$, so that the polarization vector becomes
\begin{align}
\label{eqn:rg:modl}
	 \vec{e_p} \rightarrow \begin{pmatrix}
	     \cos \! \left [ \theta + \delta \right ] \\ 
	     \sin \! \left [ \theta + \delta \right ] 
	 \end{pmatrix} \! . 
\end{align}
The random angle~$\delta$ is drawn from a narrow Gaussian distribution centered at zero.
In this way, variability in the orientation of cell polarization upon polarity bursts is taken into account.

The model reproduces all essential features of the experimentally observed dynamics of the cell-cargo motion as illustrated in Fig.~\ref{fig:4}. 
The time series of the cell-cargo distance~(Fig.~\ref{fig:4}a) shows an intermittent dynamics with pronounced, recurring downward spikes perturbing the fluctuations around the characteristic distance~$l_0$. 
Spikes correspond to polarization events during which the cell pulls the cargo from its leading to its trailing edge.  
The associated histogram~(Fig.~\ref{fig:4}b) of the cell-cargo distance reveals a pronounced peak and a characteristic asymmetry, similar to the experimentally observed histograms~(cf.~Fig.~\ref{fig:2}).
Cargo trajectories in the frame of reference of the cell extracted from simulations are approximately circular with occasional crossings due to polarity induced persistent motion~(Fig.~\ref{fig:4}c), and the corresponding recurrence plots show a checkerboard structure, consistent with the experimental results, see Fig.~\ref{fig:4}d.

\paragraph*{Model simulations of long-time dynamics predict an optimal cargo size that maximizes truck diffusivity}

\begin{figure}[b]
\centering
\includegraphics[width=\columnwidth]{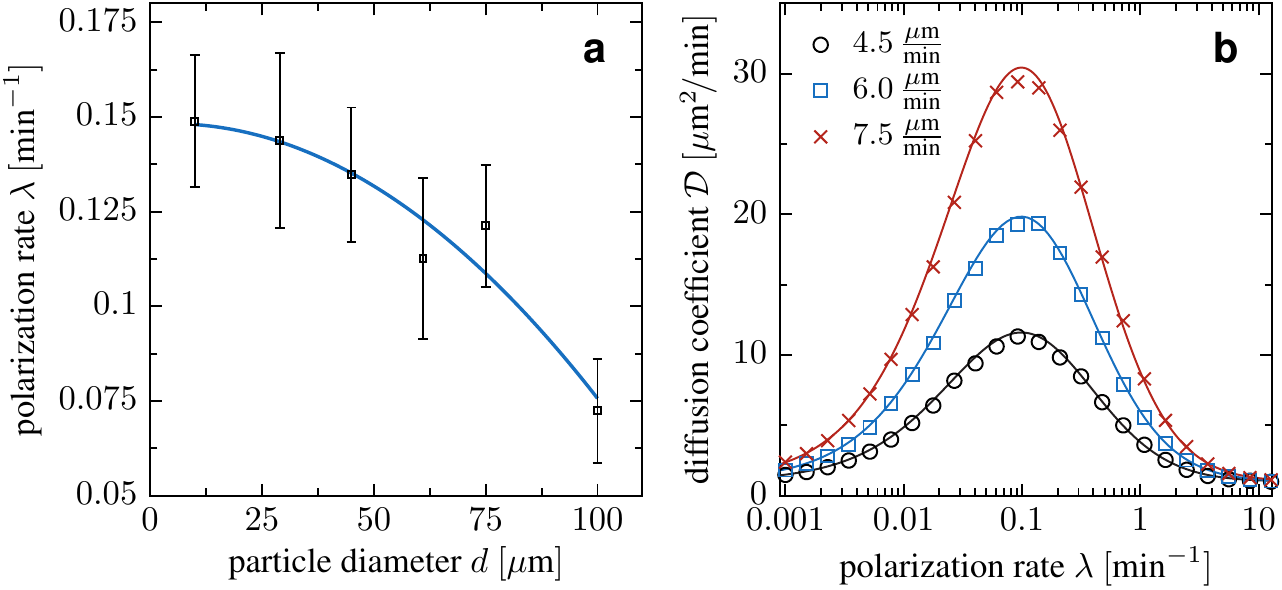}
\caption{\fighead{Long-time transport properties of the cell-cargo truck.} (\fl{a}) Dependence of the polarization rate~$\lambda$ on the particle diameter inferred from experimental data~(cf.~Methods), along with a parabolic fit indicating a monotonous decay. (\fl{b}) Non-monotonic dependence of the long-time diffusion coefficient on the polarization rate~$\lambda$ as predicted by the active particle model. Simulations~(points) and theoretical prediction~(lines), cf.~Methods, are in full agreement. The comparison is shown for three different speed values:~$v_0 = 4.5\;\! \mbox{\textmugreek m}/\mbox{min}$ (black circles), $v_0 = 6 \;\! \mbox{\textmugreek m}/\mbox{min}$ (blue squares) and $v_0 = 7.5\;\! \mbox{\textmugreek m}/\mbox{min}$ (red crosses). Other parameters are identical to Fig.~\ref{fig:4}. The monotonic decay of the polarization rate as a function of the particle diameter together with the non-monotonic scaling of the long-time diffusion coefficient of the truck as a function of the polarization rate suggest that the existence of an optimal cargo diameter which maximizes the transport capabilities of the cellular truck. } 
\label{fig:5}
\end{figure}

Polarity bursts followed by persistent motion are the key dynamical feature that drives the net motion of a cellular truck.
At the modeling level, the polarization rate~$\lambda$ is thus the central parameter that controls the long-time diffusion constant~$\mathcal{D}$ of the truck.
Experimentally, we observed that~$\lambda$ decreases with cargo diameter~$d$, see Fig.~\ref{fig:2} and Fig.~\ref{fig:5}a.
In experiments, the rate~$\lambda$ is thus indirectly controlled by choosing the cargo diameter.
Together, the two dependencies~$\lambda(d)$ and~$\mathcal{D}(\lambda)$ determine the pivotal role of cell-cargo interaction for cargo transport.

Simulations of the cellular truck model~[Eqs.~\eqref{eqn:rg:mod1}-\eqref{eqn:rg:modl}] enable us to predict the long-time dynamics of the truck as a whole, quantified based on the mean-squared displacement. 
The model predicts a non-monotonic dependence of the diffusion coefficient~$\mathcal{D}$ on the polarization rate~$\lambda$. 
This is illustrated in Fig.~\ref{fig:5}b, where the diffusion coefficient, numerically calculated for different polarization rates~$\lambda$, is plotted together with an analytical estimate of~$\mathcal{D}(\lambda)$ derived via a systematic mode reduction of the master equation corresponding to the particle-based Langevin model~(mathematical details of the derivation are provided as SM~\cite{SM}). 
Both, numerical simulations and analytical estimates predict a markedly peaked functional dependence~$\mathcal{D}(\lambda)$.
The maximum is located around values of~$\lambda$ comparable to the depolarization rate~$\sigma$, which is defined as the inverse of the lifetime of the polarized state of the cell, encoding the experimentally observed persistence in cell motility during the motile phase~(cf.~Fig.~\ref{fig:1}).
As discussed previously, most of the displacement of the cellular truck is observed during this phase. 
Since the polarization rate decreases monotonically with increasing particle diameter~$d$ and the depolarization rate~$\sigma$ is independent of~$d$ as it is inherent to the cell, the model thus predicts a non-monotonic dependence of the diffusion coefficient on cargo size.
In short, an optimal cargo size maximizes the transport efficiency of a cellular truck.

\paragraph*{Scaling arguments and experimental data confirm an optimal cargo size}

To comprehend the physical mechanism behind the non-monotonic relation between cargo size and truck diffusivity and in order to provide an intuition for the peak location, we present a simple scaling argument.
There are two relevant timescales in this setting:~first, the dwelling time~$\tau=\lambda^{-1}$ that is set by the polarization rate $\lambda$ and, second, the lifetime of the polarized state~$\tau_p=\sigma^{-1}$. 
Let us approximate the trajectory of a truck as a sequence of independent, persistent runs of length~$l_\textup{run}$, interrupted by reorientations (polarization events that occur at a rate $\lambda$).
To lowest order, we may estimate the diffusion coefficient of the truck as~$\mathcal{D}\sim l_\textup{run}^2 / \tau$, where $l_\textup{run} \approx v_0 \tau_\textup{run}$ is the typical run length between two reorientation events and $\tau_\textup{run}$ is the time for which a truck is actually running. 
The run time~$\tau_\textup{run}$ depends on the lifetime of the polarized state~$\tau_p$ and the dwelling time~$\tau$.
If the lifetime of cell polarization is shorter than the typical time at which reorientations occur, i.e.~$\tau_p \ll \tau$, the run-time equals the cell polarization time~($\tau_\textup{run} = \tau_p$) such that~$l_\textup{run} = v_0 \tau_p$ and, consequently, $\mathcal{D}\sim \tau^{-1}$. 
However, in the opposite limit~($\tau_p \gg \tau$), runs are cut short because reorientations may be spontaneously triggered during a run.
Accordingly, the run time is determined by the inverse reorientation rate~($\tau_\textup{run} = \tau$) thus implying the scaling $\mathcal{D}\sim \tau$.
Therefore, the following non-monotonic dependence of the diffusion coefficient on~$\tau$ is expected: 
\begin{equation}
\mathcal{D} \sim
\begin{cases}
\tau, &\text{$\tau \ll \tau_p$}, \\
\tau^{-1}, &\text{$\tau \gg \tau_p$}.
\end{cases}
\end{equation}
While the lifetime of cell polarization~$\tau_p$ is an intrinsic feature of the cell, the polarization rate~$\lambda$ depends on cargo size, as it decreases monotonically with the particle diameter according to our experimental findings~(Fig.~\ref{fig:5}a).
Therefore, an optimal particle size for cell-driven transport is expected, in line with the predictions of model simulations presented above:~small particles cause frequent reorientations thus stopping runs too early; on the other hand, large particles let the cell-cargo system dwell in a non-polar, non-motile state, which is inefficient for transport.
This non-monotonic scaling of the diffusion coefficient with particle size represents our main insight into cell-driven cargo transport.

The theoretical prediction of a non-monotonic trend in the transport efficiency as a function of cargo diameter also yield an explanation of our initial experimental observations displayed in Fig.~\ref{fig:1}b.
Notably, cellular trucks loaded with particles of intermediate size spread faster than unloaded cells. 
This counterintuitive observation supports our modeling hypothesis that symmetry-breaking induced by the mechanical cell-cargo interaction triggers repeated bursts of cell polarization, thereby promoting motility of the cellular truck as a whole. 
Note also that si\-mu\-la\-tions of our model returned realistic values for the diffusion coefficient~$\mathcal{D}$, predicting its maximum if polarization and depolarization rate are comparable~($\lambda \simeq \sigma$).

\paragraph*{Conclusion}

In this study, we highlighted the potential of motile amoeboid cells to act as autonomous carriers of micron-sized cargo and investigated the transport dynamics of this novel biohybrid system. 
Our experiments revealed a complex cell-cargo motion pattern involving recurrent phases of polar displacement that alternate with an isotropic resting state. 
This can be qualitatively explained based on the mechano-responsiveness of amoeboid cells:~their crawling motility relies on cyclic cell shape changes to pull the cell body forward, a process that is governed by surface adhesion in conjunction with the mechanical stresses generated by the cell's cytoskeletal machinery~\cite{ananthakrishnan2007forces,alvarez2015three,copos2017mechanosensitive}.
The geometry of the surrounding environment, such as the presence of a cargo, thus strongly affects cell polarity and, consequently, migration patterns~\cite{petrie2009random,nagel2014geometry,doyle2009one,boneschansker2014microfluidic,pieuchot2018curvotaxis}. 
In our specific case, the cell is positioned in a confined space between the coverslip and cargo surfaces, where the latter provides an additional substrate for adhesion compared to a cell without cargo.
This may explain the spontaneous bursts of cell polarity observed in our experiments, as adhesion to the cargo triggers additional actin activity, which in turn induces cytoskeletal symmetry breaking, thus promoting motility along the direction of the cell-cargo axis.
The characteristic lifetime of cell polarity~$\tau_p = \sigma^{-1}$ results in persistent motion~\cite{gorelik2014quantitative,petrie2009random,prentice2016directional}, causing the cell to keep moving in the same direction.
As a consequence, the cell pulls the cargo forward, resulting in a net displacement during these motile phases. 
In short, confinement-induced cell-cargo interactions determine the transport capacities as a whole. 
We therefore envision that different cargo shapes and sizes may significantly impact the cytoskeletal activity and polarization dynamics, allowing for a dedicated design of the transport properties of active biohybrids in the future.

The theoretical framework that we developed here connects the experimentally observed cell-cargo dynamics at the level of the individual cellular truck with the long-time displacement of the biohybrid system as a whole, thereby enabling us to predict the effective transport properties.
In particular, the theoretical analysis unveiled how the transport efficiency is determined by the interplay between the intrinsic persistence of cell polarity and characteristic time scales of the cell-cargo interaction, such as the polarization rate.
Notably, we predict that an optimal cargo size enhances the diffusivity of cellular trucks, such that loaded cells may even outperform unloaded ones in terms of their long-time transport capacity; a result that found confirmation in our experiments.


\small{

\section*{Appendix:~Methods}

\vspace{-0.35cm}

\metparagraph*{Cell culturing.}

LimE-mRFP MyoII-GFP expressing AX2 {\dicty} cells (kindly provided by G\"unther Gerisch, Martinsried) are adherently grown on the bottom surface of tissue culture flasks (TC Flask T75 Standard, Sarstedt AG \& Co.~KG, N\"umbrecht, Germany), cultured at \SI{22}{\degreeCelsius} in nutrient medium (HL5 medium including glucose, Formedium Ltd., Norfolk, England) supplemented with $1\%$ penicillin/streptomycin antibiotics mix~(CELLPURE{\tiny \textregistered} Pen/Strep-PreMix, Carl Roth GmbH+Co.~KG, Karlsruhe, Germany). 
In addition, $0.05\%$ of G418 (G418 disulfate ultrapure, VWR International, LLC.) and $0.1\%$ of Blasticidin (Blasticidin S (hydrochloride), \textsc{Cayman chemical}, Ann Arbor, Michigan, USA) were added as selection agents.
To avoid confluency, cells were subcultured into new flasks every two days via 1:20 dilution of a cell suspension from a previous flask, or 1:64 prior to weekends.
Cell cultures have been renewed every four weeks to avoid accumulation of any undesired mutation eventually arising by genetic drift. 
In addition, we also used LifeAct-mRFP expressing AX2 {\dicty} cells (kindly provided by Maja Marinowic and Igor Weger, Zagreb), which were cultivated analogously, the only difference being the use of just one selection agent~(G418).

\metparagraph*{Sample preparation.}

Polystyrene spherical microparticles~(Polybead{\tiny \textregistered} Microspheres, Polysciences Europe GmbH, Hirschberg an der Bergstrasse, Germany and Polystyrene particles (PS-R), microParticles GmbH, Berlin, Germany) are suspended in phosphate buffer (S\o rensen buffer: in dd\ce{H2O}, \SI{2}{\gram\per\litre} \ce{KH2PO4}, \SI{0.36}{\gram\per\litre} \ce{Na2HPO4}, \SI{50}{\micro\litre\per\litre} 1M \ce{MgCl2}, \SI{50}{\micro\litre\per\litre} 1M \ce{CaCl2}; pH 6.0) to have a particle number density of $\SI{5d5}{\per\milli\litre}$.

Cells are harvested from a flask, normally during the subculturing procedure, always far from confluency.
The obtained cell suspension is then diluted to obtain a cell count of roughly $\SI{50d3}{\per\milli\litre}$, thus \SI{2}{\milli\litre} of the new suspension are put into a culture dish (FluoroDish\textsuperscript{TM} tissue culture dish with cover glass bottom - 35mm, 23mm well, World Precision Instruments, Inc., Sarasota, Florida, USA).
The sample is left settling for \SI{15}{\minute} to let cells sediment and adhere on the bottom of a dish; afterwards, \SI{50}{\micro\litre} of particle suspension are added while gently shaking the dish in order to achieve a more uniform particle distribution on the substrate.

\metparagraph*{Imaging.}

The sample is imaged in time-lapse by means of multi-channel confocal imaging using a Laser Scanning Microscope~(LSM 780, Zeiss, Oberkochen, Germany).
A DPSS laser, generating $\SI{561}{\nano\metre}$ electromagnetic radiation, excites the fluorophore mRFP co-localizing with F-Actin, whose emission is band-pass filtered ($582-\SI{754}{\nano\metre}$) and then detected by a photomultiplier.
An argon-ion laser, emitting at the wavelength~$\SI{488}{\nano\metre}$, is used to excite the fluorophore GFP co-localizing with myosin II; again, the fluorescence emission is band-pass filtered ($493-\SI{556}{\nano\metre}$) before detection.
All the transmitted light is collected by a third acquisition channel, used for particle imaging, where images arise from the contrast generated by discontinuities of the refractive index.

The focal plane is set to be in correspondence with the substrate surface to focus the ventral surface of cells together with a bottom section of the particle.
The depth of field has been roughly \SI{0.7}{\micro\metre} to \SI{0.8}{\micro\metre} (detector pinhole aperture of $1\,$Airy Units, 63x/40x objective), enabling to collect light from an optimal volume of the cell body, granting good signal-to-noise ratio at low laser intensities without excessively compromising the resolution of the optical sectioning.
For any detected cell-particle system, images are recorded every $\SI{10}{\second}$, until the system of interest leaves the field of view or becomes unsuitable~(due to cell division or other cells/particles alter the configuration of the system), for a maximum time of $1\!\:$h.

\metparagraph*{Image analysis.}

The image processing and data analysis is performed using custom algorithms written as \textsc{Matlab} (MathWorks{\tiny \textregistered} Inc.) code.

Cell segmentation relies on the $\SI{561}{\nano\metre}$-source fluorescence channel, exploiting the contrast created by the fluorescence emission from the mRFP labeling the F-actin, while particle segmentation is based on brightfield images from the transmitted light channel.
For both objects, images first undergo median filtering, followed by a contrast enhancement protocol, involving a sequence of nonlinear histogram remapping steps. 
The processed images are then binarized with a threshold determined by Otsu's method~\cite{otsu1979threshold} and the segmented objects are tracked through the frames.
In the case of cells, boundaries detected from segmentation are further processed with an active contouring algorithm~\cite{xu1998snakes}. 
The two-dimensional position vectors~$\vec{r}$ and~$\vec{R}$ for cell and cargo, respectively, defined in the focal plane parallel to the substrate surface, are defined as the centroids of the objects that were identified by binarization.

\metparagraph*{Data analysis.}

To compensate for the cell-to-cell variability in size~--~the cell size sets the spatial scale of relative motion of cell and cargo and, thus, affects relative distance distributions~--~the cell-cargo distance~$l(t) = \abs{\vec{R} - \vec{r}}$ was rescaled by an effective cell radius~$r_{c}$, that was determined to match the average projected cell area~$\mean{A_{c}} = \pi r_{c}^2$, prior to the calculation of the population-averaged histograms of the cell-cargo distance as shown in Fig.~\ref{fig:3}.

For the experimental estimation of the polarization rate~$\lambda$, detection of polarization events was performed as follows. 
First, the time series~$l(t) = \abs{\vec{R}(t)-\vec{r}(t)}$ of the cell-cargo distance was renormalized by subtracting the time-averaged mean~$\overline{l(t)}$ and subsequent division by the standard deviation in each recording.
Population-averaged histograms of all observations revealed a peaked, asymmetric distribution, comparable to Fig.~\ref{fig:4}b, which is well-approximated by the sum of two Gaussian distributions. 
The main contribution of the histogram stems from resting and motile phases, where $l(t)$ fluctuates around a preferred value~$l_0$; the asymmetry stems from transitions during which the distance~$l(t)$ becomes small. 
The inspection of all histograms suggested that fluctuations of the distance $l(t)$ towards zero by more than $1.75$-standard deviations from its mean value indicates a transition and therefore provides a reliable cut-off criterion. 
Based on this cut-off, the time series was binarized and transitions were counted. 
The estimation of the polarization rate~$\lambda$ from the detected events is based on the assumption that the polarization process is Poissonian. 
Accordingly, the probability to observe~$k$ polarization events in a time interval~$\tau$ is given by the Poisson distribution
\begin{align}
    P_{\tau}(k|\lambda) = \frac{(\lambda \tau)^k e^{-\lambda\tau}}{k!}, 
\end{align}
implying the likelihood
\begin{align}
    \mathcal{L} = \prod_{i=1}^N P_{\tau_i}(k_i|\lambda). 
\end{align}
For each colloid size, we observed~$N=10$ independent experimental realizations of length~$\tau_i$ and counted the number of polarization events~$k_i$~($i=1,2,...,N$). 
The maximum-likelihood estimator~$\hat{\lambda}$ for the rate~$\lambda$ is determined by the total number of observed events divided by the total observation time: 
\begin{align}
    \hat{\lambda} = \frac{ \sum_{i=1}^N k_i }{\sum_{j=1}^N \tau_j}. 
\end{align}
In the vicinity of its maximum value, we approximate the likelihood as a Gaussian distribution with mean~$\hat{\lambda}$ and the standard deviation
\begin{align}
    \sigma_{\lambda} = \frac{\hat{\lambda}}{\sqrt{\sum_{j=1}^N k_j}} .
\end{align}
In Fig.~\ref{fig:5}a, the error bars graphically represent this~$1\sigma$-interval. 
This automatic transition counting yields consistent results with manual counting of transitions which is based on the binarizing and thresholding of recurrence plots of the distance $l(t)$ via Otsu's method, where such event appears as a distinct horizontal/vertical lines.

The temporal dependence of the mean-squared displacement of colloids, as shown in Fig.~\ref{fig:1}b, was estimated as follows. 
At first, the time-averaged mean-squared displacement
\begin{align}
    \label{eqn:tamsd}
    \delta_{\mu} \! \left ( m \cdot \Delta \right) = \frac{1}{n_{\mu} - m} \sum_{k=1}^{n_{\mu} - m} \Big | \vec{R}_{\mu}(t_{k+m}) - \vec{R}_{\mu} (t_{k}) \Big |^2
\end{align}
was calculated for each trajectory, where~$\vec{R}_{\mu}(t_{k})$ denotes the position of colloid~$\mu$ in frame~$k$, $n_{\mu}$ is the total number of frames in the $\mu$-th trajectory and $\Delta$ is the time step of image acquisition~(inverse frame rate). 
We assume that the colloid displacements are Gaussian distributed with zero mean in two dimensions,
\begin{align}
    \bar{P} (\vec{R} - \vec{R}_0) = \frac{1}{\pi l_2} \exp \! \left ( - \frac{\abs{\vec{R} - \vec{R}_0}^2}{l_2} \right )
\end{align}
which is justified at least for long time intervals. 
The aim is to estimate the width of this Gaussian distribution as a function of the lag parameter~$m$ which is the ensemble averaged mean-square displacement~$l_2$. 
Given the measured colloid displacements are Gaussian and independent, the time averaged mean-squared displacement~$\delta_{\mu} \! \left ( m \cdot \Delta \right)$ of a single trajectory follows a Gamma distribution,
\begin{align}
    p(\delta) = \frac{\delta^{\alpha-1}}{\Gamma(\alpha)} \beta^{\alpha} e^{-\beta \delta}
\end{align}
with a shape parameter $\alpha = n_{\mu} - m$ and the inverse scale parameter~$\beta = (n_{\mu} - m)/l_2$. 
From this distribution, we construct a likelihood for the observed displacements, given the parameter~$l_2$ to be estimated as a function of the time lag~$m\cdot \Delta$. 
For small time lags~--~if the time lag is shorter than the shortest trajectory~--~the maximum likelihood estimator for the ensemble-averaged mean-squared displacements equals the weighted average of the time-averaged mean-squared displacements of all trajectories,
\begin{align}
    \label{eqn:weightmsd}
    l_2 \! \left ( m \cdot \Delta \right ) = \frac{\sum_{\mu=1}^N (n_{\mu}-m) \delta_{\mu} \! \left ( m \cdot \Delta \right) }{ \sum_{\nu=1}^N (n_{\nu}-m) }     ,     
\end{align}
where the weights correspond to the number of terms in the sum of Eq.~\eqref{eqn:tamsd}, i.e.~the number of displacements they are calculated from, and~$N$ is the total number of tracks. 
As the time lag increases, it may become larger than the number of frames in short tracks and, consequently, Eq.~\eqref{eqn:weightmsd} looses its applicability. 
In that case, we construct the likelihood as a product as follows
\begin{align}
    \label{eqn:likelihood}
    \mathcal{L} = \left [ \prod_{\mu=1}^{N_c} p \big ( \delta_{\mu} \! \left ( m \cdot \Delta \right) \! \big ) \right] \! \cdot \! \left [ \prod_{\nu=1}^{N_i} P \big ( \delta_{\nu} \! \left ( n_{\nu} \cdot \Delta \right) \big ) \! \right]
\end{align}
where the first product runs over the~$N_c$ trajectories which contain more than~$m$ frames and the second product includes those $N_i$ tracks which have less or equal than~$m$ frames ($N = N_c + N_i$). 
In the second product, 
\begin{align}
    P(\delta) = \int_{\delta}^{\infty} d\delta' \, p \! \left ( \delta' \right )
\end{align}
is the probability that the time-averaged mean-squared displacement is larger than~$\delta$. 
In this way, the information is taken into account that the corresponding colloid was observed to displace by at least~$\delta_{\mu} \! \left ( n_{\mu} \Delta \right )$ for large time lags. 
This construction of the likelihood is based on the assumption that the mean-squared displacement is non-decreasing function in the long-time limit. 
We eventually maximized the likelihood~[Eq.~\eqref{eqn:likelihood}] numerically to obtain an estimator for~$l_2$.

\metparagraph*{Theoretical estimation of the diffusion coefficient.}

In order to assess how the long-time transport properties of the cell-cargo truck depend on the parameter values, we derived the long-time limit analytically. 
For this purpose, the cell-cargo model dynamics is simplified as follows: (i) in the isotropic state, the cell undergoes Brownian diffusion while the cargo is located at a fixed distance from the cell, performing random motion around it; (ii) in the polarized state, the cell performs a persistent random walk with a fixed speed~$v_0$ and a finite persistence length parametrized by the angular noise strength~$D_{\varphi}$, carrying the cargo on its back; (iii)~transition phases~--~polarity bursts during which the cell pulls the cargo from the leading to the trailing edge~--~are fast processes compared to other relevant timescales and are therefore assumed to occur instantaneously. 
This corresponds, formally, to the limit~$\alpha \rightarrow \infty$, implying that we neglect the timescale of the relative dynamics of cell and cargo. 
The simplified model is formally represented by two coupled master equations.
We performed an expansion in Fourier modes. 
Only one Fourier coefficient, corresponding to the density of cellular trucks~$\rho(\vec{r},t)$ at position~$\vec{r}$ at time~$t$, is a slow variable~(conserved quantity) that dominates the long-time dynamics.
Therefore, the coupled system of Fourier modes can systematically be reduced onto the density by adiabatic elimination of fast modes. 
To lowest order in spatial gradients, the diffusion equation is obtained from which the diffusion constant~$\mathcal{D}$ can be read off. 
Mathematical details of the calculation are provided as SM~\cite{SM}; see also Ref.~\cite{nava_markovian_2018} where similar methods were applied for coarse-graining.

The above mentioned simplified model overestimates the actual diffusion coefficient as changes of the cargo position during polarity bursts of the cell are not instantaneous. 
During polarity bursts, during which the cell and particle interchange their positions, little net displacement of the cellular truck as such occurs. 
We developed a heuristic correction based on the idea that a run of the cell can be subdivided into two phases:~first, the polarized cell crawls underneath the particle and pulls it onto its back and, second, it moves persistently with the cell on its back (motile phase), cf.~Fig.~\ref{fig:1}. 
We assume that no active transport occurs as cell and cargo change their relative positions.
Accordingly, the actual time spent in the persistent run phase is decreased. 
This leads to a correction factor $\alpha / (\alpha + \lambda)$ of the diffusion coefficient which equals the probability not to be in the resting phase.

}


\section*{Acknowledgments}

V.L., S.K., R.L. and C.B.~acknowledge financial support via the IMPRS \textit{Multiscale Bio-Systems}, S.S.P.~and C.B.~thank the Deutsche Forschungsgemeinschaft (DFG) for funding (Sachbeihilfe BE~3978/3-3).  
We thank Kirsten Sachse and Maike Stange for technical support and Fernando Peruani for valuable comments on the manuscript.

\section*{Author contributions}

V.L.~performed experimental research, V.L., R.G.~analyzed data and designed modeling framework, R.G.~performed analytical derivations and mathematical analysis, O.N., S.S.P.~contributed experimental data, V.L., R.G., C.B.~wrote the manuscript, S.K., R.L.~co-supervised the project, C.B.~designed research and supervised the project.

\section*{Competing interests}

The authors declare no competing interests.

\section*{Data availability}

The data that support the plots within this paper and other findings of this study are available from the corresponding author upon request.

\vspace{0.5em}

\noindent
\textbf{Correspondence and requests for materials} should be addressed to C.B.



%

\end{document}